\documentclass[useAMS,usenatbib,usegraphicx]{mn2e}
\def\eqq#1{Equation~(\ref{#1})}
\newcommand\etal{{\it et al.\/}}
\newcommand\bfe{\mbox{\boldmath $\eta$}}

\newcommand{\mnras}{MNRAS}
\newcommand{\aj}{AJ}
\newcommand{\apj}{ApJ}

\newcommand{\aap}{AAP}

\title[Shape measurement biases]{Shape measurement biases from underfitting and ellipticity gradients}
\author[Bernstein]{Gary M. Bernstein\thanks{garyb@physics.upenn.edu } \\
 Dept. of Physics and Astronomy, University of Pennsylvania}

\begin{document}
\maketitle

\begin{abstract}
Precision weak gravitational lensing experiments require measurements of galaxy shapes accurate to $<1$ part in 1000.  We investigate measurement biases, noted by
Voigt and Bridle (2009) and Melchior {\it et al.} (2009), that are common to shape measurement methodologies that rely upon fitting elliptical-isophote galaxy models to observed data.  The first bias arises when the true galaxy shapes do not match the models being fit.  We show that this ``underfitting bias'' is due, at root, to these methods' attempts to use information at high spatial frequencies that has been destroyed by the convolution with the point-spread function (PSF) and/or by sampling.  We propose a new shape-measurement technique that is explicitly confined to observable regions of $k$-space.  A second bias arises for galaxies whose ellipticity varies with radius.  For most shape-measurement methods, such galaxies are subject to ``ellipticity gradient bias.'' We show how to reduce such biases by factors of 20--100 within the new shape-measurement method.  The resulting shear estimator has multiplicative errors $<1$ part in $10^3$ for high-$S/N$ images, even for highly asymmetric galaxies.  Without any training or recalibration, the new method obtains $Q=3000$ in the GREAT08 Challenge of blind shear reconstruction on low-noise galaxies, several times better than any previous method.
\end{abstract}


\section{Introduction}
The appearance of background galaxies are subtly distorted due to weak gravitational lensing (WL) by mass along the line of sight.  This WL effect is detectable as a breaking of symmetries in the intrinsic background scene, in particular a shearing that breaks the isotropy of the galaxy ellipticity distribution.  Because this ``cosmic shear'' is a direct measure of the metric fluctuations in the Universe, it is readily used to constrain the dark matter and energy that dominate the gravitational behavior of our Universe.  A review of the cosmic shear method is provided by \citet{HoekstraJain} and the role of WL in the context of dark energy studies is summarized in \citet{DETF}.

The cosmic shear signal is subtle and difficult to measure to high accuracy.  The shear along a typical line of sight on cosmological scales causes a $\approx 2\%$ change in axis ratio of a background source.
Propagation of shear-estimation errors into cosmological parameter inferences suggests that systematic errors in shear estimation must be held below $\approx 1$ part in 1000 of the expected shear to avoid significant degradation of cosmological inferences \citep{HTBJ, AmaraRefregier}.  Roughly speaking this means that systematic errors in the ellipticity $e$ of individual galaxies must be held below 1 part in 1000, {\it i.e.} $\langle\delta e\rangle/e \la 10^{-3}$.   In community-wide blind tests of shear recovery from simulated data \citep{STEP1, GR08}, no methods have yet achieved this level of accuracy, even for simulated simplified images with unrealistically high signal-to-noise ($S/N$) levels and perfect knowledge of the point spread function (PSF).  As most recently reported by \citet{GR08}, methods that could be practically applied to real survey data produced shear estimates that were in error by $\approx 1\%$ or more of the typical cosmic shear signal.

A biased and/or noisy shape measurement method can nonetheless be applied to real data if its biases are calibrated and compensated.  A simple scheme is to fit the measured shear $\gamma_{\rm meas}$ to the true (simulated) shear $\gamma_{\rm true}$ via
\begin{equation}
\gamma_{\rm meas} =(1+ m) \gamma_{\rm true} + c.
\end{equation}
The multiplicative and additive error terms $m$ and $c$ derived from simulations could then be applied to measured data.\footnote{We write equations treating $\gamma$ as a scalar, for simplicity.  In practice the shear, ellipticities, additive and multiplicative error coefficients all have two components.  We will also ignore the distinction between the shear $\gamma$  and the reduced shear $g=\gamma/(1-\kappa)$.}
The difficulty is that $m$ and $c$ will depend upon sets of parameters $\pi_g$ and $\pi_\star$ describing the properties of the source galaxies and point-spread function (PSF), respectively.  If the simulation does not properly mimic the real populations, then the empirical correction will be in error---and we may not even know which parameters are the critical ones to emulate properly.  A Bayesian formalism can be used to derive an unbiased shear estimator $\hat\gamma$ from a collection of potentially biased shape estimates $\hat e_i$ {\em if} we know all three of: the intrinsic distribution of galaxy shapes $P(e)$; the change $dP(e)/d\gamma$ of the shape distribution under applied shear; and the conditional probability $P(\hat e | e)$ of the measurement process \citep{lensfit}.  The need for such knowledge was previously noted by \citet[BJ02]{BJ02}.  The calibration is difficult, however, because the measurement process noise and bias will depend upon galaxy and PSF properties, $P(\hat e | e, \pi_g, \pi_\star)$, so we need to know this function and also the distribution $P(e, \pi_g)$ of all galaxy subsets of interest.  Characterizing all of these functions by simulation will be hard because of the wide variety of galaxy shapes on the sky and PSF variation, especially if we do not know exactly what parameters are important.  

It is therefore of great interest to produce a shape-measurement algorithm that is minimally biased, and which produces distributions $P(\hat e | e)$ that are robust, {\it i.e.} characterized by a few well understood properties of the galaxy or PSF.  Any corrections that must be derived empirically (via simulation) should be as small as possible to minimize errors that arise from differences between the simulations and the real observations.

Furthermore the calibration depends on knowing the response $dP(e)/d\gamma$ of the noise-free, unsheared ellipticity population $P(e)$ to an applied shear.  If the shape-measurement methodology misestimates $dP/d\gamma$ even in the absence of noise, then the calibration will be in error even if $P(\hat e | e)$ is known.  If $dP/d\gamma$ depends upon the detailed parameters $\pi_g$ and/or $\pi_\star$, then the calibration will depend upon knowledge of the complete distribution $P(e,\pi_g, \pi_\star)$ at all points in the survey.

In this context it is worrisome that \citet[VB]{VB} and \citet[MBLB]{MBLB} demonstrate that shape-measurement techniques based on model-fitting to the observed data can be biased by $\ga 1$ part in 100 even in the absence of measurement noise.  These biases arise when the models being fit are not well matched to the galaxies being measured, for example if an exponential-disk model or Gauss-Laguerre expansion is being fit to a deVaucouleurs profile galaxy.  In principle the shape measurement bias can be reduced by adding degrees of freedom to the model so that it spans all possible galaxy shapes, for example by using a complete set of basis functions to model the galaxy.  This will, however, come at the expense of increased uncertainty on the inferred shear as the model fits become increasingly degenerate.

In this paper we examine these ``underfitting'' biases more closely, and devise a shape-measurement technique that avoids the issue.  Then we investigate a more general
effect revealed by the VB tests, in which a galaxy that has ellipticity varying with radius (or more precisely, with the scale of the weight function) leads to a mis-estimation of the applied lensing shear.  We will show how this ``ellipticity gradient bias'' can be quantified and corrected using our newly proposed shear-measurement method.  After controlling these two biases we are able to measure shear from high-$S/N$ galaxies to the desired accuracy.

\section{Roundness-test methods}
BJ02 introduced a geometric definition of galaxy shape which is intended to eliminate the need for empirical shear calibration factors, at least for high-$S/N$ objects.  This shape definition renders the derivative $dP(e)/d\gamma$ a universal function of $P(e)$, independent of any galaxy or PSF parameters.  We briefly summarize the notion of ``geometric'' shape assignment.  

The shearing action of lensing that we seek to detect is a mapping of  the unlensed galaxy appearance $I_u$ to the apparent surface brightness $I_a({\bf x})$ via $I_a({\bf x}) = I_u({\bf S}^{-1}_\eta{\bf x}).$  We take a unit-determinant version of the shear, so that for a shear along the $x$ axis, 
\begin{equation}
{\bf S}_\eta = \left( \begin{array}{cc}
e^{\eta/2} & 0 \\
0 & e^{-\eta/2}
\end{array}
\right).
\end{equation}
For shear at an angle $\beta$ to the $x$ axis, we apply the appropriate rotation matrices to ${\bf S}$ and
define $\eta_1=\eta\cos 2\beta,$ $\eta_2=\eta\sin 2\beta.$  It is more common to write a traceless magnification matrix
\begin{equation}
{\bf S}_\gamma = \left( \begin{array}{cc}
1+\gamma_1 & \gamma_2 \\
\gamma_2 & 1-\gamma_1
\end{array}
\right).
\end{equation}
The two variables for shear are related by $\gamma = \tanh \eta/2$.
A shear-measurement method must infer the components of the applied shear $\eta$ from a collection $e_i$ of shapes assigned to galaxies $i\in{1,\ldots,N}$.  One crucial question is: how do we define $e_i$ for a galaxy with some $I_a({\bf x})$?

We will call a galaxy ``perfectly elliptical'' if
all of its isophotes are true ellipses with constant axis ratio $a/b$ and position angle $\beta$.  For perfectly elliptical galaxies, the galaxy $e$ is straightforwardly defined as the isophotal ellipticity $(a^2-b^2)/(a^2+b^2)$.  Furthermore an applied shear $\eta$ maps every such galaxy into another perfectly elliptical galaxy with ellipticity $e^\prime$, and the map $e^\prime(e,\eta)$ has the desired property of being independent of all other galaxy parameters.  For an applied shear $\eta\ll 1$ and an unlensed galaxy ellipticity oriented along the $x$ axis, the map can be linearized to (BJ02):
\begin{equation}
\label{etransform}
\begin{array}{ccl}
e^\prime_1 & = & e + (1-e^2) \eta \cos 2\beta \\
e^\prime_2 & = & \eta \sin 2\beta 
\end{array}
\end{equation}
Hence if all galaxies were perfectly elliptical and it were possible to observe their pre-seeing shapes $I_a(\bf x)$ at infinite $S/N$, then a perfect shear estimator would arise from simply measuring the ellipticity of any single isophote.  

Among the many problems with the above method is that galaxies do not have perfectly similar elliptical isophotes.  BJ02 show that one can assign an ellipticity $e$ to a generic galaxy $I_a({\bf x})$ such that it is transformed by shear under the same model-independent formulae as the perfect ellipticities above.  To do so we define a pair of {\em roundness tests} $t_1(I_a), t_2(I_a)$ for which:
\begin{enumerate}
\item There is one and only one shear $\eta_t$ for which
\begin{equation}
t_1[ I_a({\bf S}_{\eta_t}{\bf x}) ] = t_2[ I_a({\bf S}_{\eta_t}{\bf x}) ] = 0.
\end{equation}
\item If $t_1(I_a)=t_2(I_a)=0$ holds for some $I_a$, then it also holds after any rotation $R_\theta$ is applied to $I_a$.
\item The form of the roundness test is invariant under shear of the target galaxy.
\end{enumerate}
The first two conditions mean that a galaxy which nulls $t_1$ and $t_2$ can be considered ``round'' in the sense that any rotated version of the galaxy is still round, and any sheared version of the galaxy is no longer round.  If a chosen galaxy nulls $t_1$ and $t_2$ at test shear $\eta_t$, we assign the galaxy the ellipticity $e = \tanh(\eta_t)$, since it is a ``round'' galaxy sheared by $\eta_t$.  The third condition then guarantees that this assigned ellipticity will transform under shear as in \eqq{etransform}.

It is often desirable for the form of $t_i$ to depend upon properties $\pi_g$ and $\pi_\star$ of the galaxy and PSF, {\it e.g.} adjusted for galaxy and PSF sizes to improve the measurement $S/N$ level.  We will emphasize later however that it is critical that the roundness test be the same for sheared and unsheared versions of the same galaxy.
We will assume that the roundness test is a linear function of $I_a({\bf x})$, and hence is also a linear function of the Fourier-domain intensity $\tilde I_a({\bf k})$.
It is typical to adopt a roundness test that has a quadrupole form, since shear induces a quadrupole on a circular galaxy.  A radial weight function $w$ remains free:
\begin{eqnarray}
t_1(I_a) & \equiv & \int dx\,dy\, I_a({\bf S}_\eta{\bf x}) w(x^2+y^2) (x^2-y^2) \\
t_2(I_a) & \equiv & \int dx\,dy\, I_a({\bf S}_\eta{\bf x}) w(x^2+y^2) (2xy) \nonumber
\end{eqnarray}

The result of the roundness tests will usually depend upon the choice of origin for the coordinate system---although there are methods for which this is not true, which work on the autocorrelation function or power spectrum of $I_a$ \citep{vWautocorr,HB09,ZhangFourier}.  To fix the coordinate origin ${\bf x}_0$  we can create two additional null tests.  These tests must satisfy the three conditions given for the shear null test, with the difference that
they are null for a unique translation ${\bf x}_0$ rather than a unique shear.  The obvious choices for linear centroid null tests are dipole moments with some radial weight $w_x$:
\begin{eqnarray}
t_x(I_a) & \equiv & \int dx\,dy\,I_a[{\bf S}_\eta({\bf x}-{\bf x}_0)] w_x(x^2+y^2) x = 0,\\
t_y(I_a) & \equiv & \int dx\,dy\, I_a[{\bf S}_\eta({\bf x}-{\bf x}_0)] w_x(x^2+y^2) y = 0. \nonumber
\end{eqnarray}
The centroid and ellipticity are hence determined by searching the $\{x_0,y_0,\eta_1,\eta_2\}$ space for the point where $t_x, t_y, t_1,$ and $t_2$ are all nulled.  When all null tests are linear functions of $I_a$, the process can be equivalently cast in either real or Fourier space.

\subsection{PSF correction with basis-set methods}
The {\em observed} galaxy image $I_o({\bf x}) = I_a({\bf x}) \otimes T({\bf x})$ is always the lensed galaxy image convolved with some instrumental transfer function (PSF) $T$.  We must however access the pre-seeing function $I_a$ somehow to apply the roundness tests. The technique proposed by BJ02 is to assume that the roundness (and centroid) tests are all weighted integrals over $I_a$ with weight functions that are members of a set $\psi_i({\bf x})$ of functions that form a complete basis for functions on the plane.  If the $\psi_i$ are orthonormal and complete, then there is a unique set of coefficients $b_i$ such that
\begin{eqnarray}
I_a({\bf S}_\eta{\bf x}) & = & \sum_i b_i \psi_i({\bf x}) \\
\Rightarrow b_i &= & \int d^2x\, I_a({\bf S}_\eta{\bf x}) \psi_i({\bf x}); \\
I_o({\bf x}) & = & \sum_i b_i \left[ \psi_i({\bf S}_\eta^{-1}{\bf x}) \otimes T({\bf x}) \right].
\end{eqnarray}
The assignment of galaxy shape $e$ thus proceeds as follows: 
\begin{enumerate}
\item For some trial $\eta$ (and centroid choice), shift and shear the basis functions $\psi_i$ and convolve them with the PSF. 
\item  Using the last of the equations above, execute a linear least-squares fit to the observed galaxy to determine the coefficients $b_i$.  In the absence of noise, the $b_i$ that provide the decomposition in the first line must be a solution of this fit with $\chi^2=0$, hence a best fit.
\item Iterate the trial shear and centroid to null the four $b_i$ which embody the roundness and centroid tests.  From the second equation we see that these $b_i$ would be the result of applying the centroid test to the pre-seeing image $I_a$.  The galaxy is assigned an ellipticity corresponding to the shear that nulls the tests. [More generally the null tests are four functions of the $b_i$, but the principle remains similar.]
\end{enumerate}

An obvious drawback to this method is that the basis set must be infinite to span the full variety of all the galaxies on the sky.  Real data will not yield a unique solution to an infinite coefficient vector $\{b_i\}$ because the data are sampled on a bounded region.  Furthermore there must be regions where the transform $\tilde T({\bf k})$ of the PSF is identically zero, which will induce degeneracies into the fit of an infinite basis set, even when there is no noise in the observations.  We must therefore apply the method with {\em a finite, incomplete basis set.}  \citet{LewisStack} notes that fitting methods can become biased estimators of shear when the models do not fit the galaxies exactly.  Here we explore the conditions under which biases develop.

\subsection{Bias from incomplete bases}
\begin{figure}
\includegraphics[width=0.99\columnwidth]{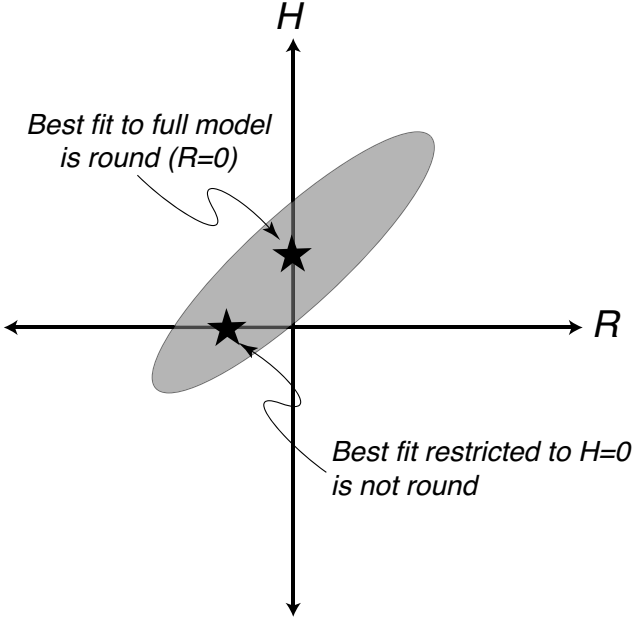}
\caption[]{Illustration of bias induced by underfitting: in this simplified picture, we plot the contours of $\chi^2$ for a fit to a galaxy with a model having two degrees of freedom, $R$ and $H$.  $R$ is a ``roundness test'', {\it i.e.} an indicator of shape, while $H$ is some additional parameter needed to obtain a complete model of the galaxy.  The true values $R_{\rm true}=0$ and $H_{\rm true}\ne 0$ are obtained from a full fit.  However if we truncate the model so that $H$ is fixed to zero, then the best fit moves to $R\ne0$, producing a bias in the measured shape.  Fitting to the full $(R,H)$ model may be disfavored, or even impossible to execute, because the resultant uncertainty on $R$ can become larger.  However the truncated model will be biased.}
\label{covar}
\end{figure}

We simplify the situation by assuming that there is one coefficient $R=b_i$ that serves as a roundness test, and another coefficient $H=b_j$ that we will consider ``hidden'' because we will later truncate it from the model.  Assume that the galaxy is round in some coordinate system, which means that a complete decomposition of $I_a$ would yield values $R_{\rm true}=0$ and $H_{\rm true}$.  As illustrated in Figure~\ref{covar}, the the fitting process may induce non-zero correlation between $R$ and $H$.

Next consider that truncating the fit is equivalent to setting $H=0$ and seeking the best fit for $R$.  If ${\rm Cov}(R,H)\ne0$ and $H_{\rm true}\ne 0$, then the best-fit value $\hat R$ will be non-zero.  {\em This will be manifested as an error in the assigned ellipticity.}  Note that the uncertainty on $R$ will be larger when $H$ is free than when $H$ is truncated; in cases with finite noise, this may eventually produce some upper limit to the number of basis functions that can be included in the model.  Hence when the measurement process correlates the roundness test with other members of the basis set, there is an inevitable tradeoff between accuracy and bias in this fitting method. 

In an ideal case of basis-function fitting, (1) the noise in the image is white, (2) there is no PSF, and (3) the data is available over the full plane.  In this case the coefficients of an orthonormal basis will be uncorrelated, so ${\rm Cov}(H,R)=0$ and there will be no bias from fitting a truncated basis set.  Correlations between basis coefficients are induced, however, by finite sampling, bounded data range, and especially by convolution with the PSF \citep[BJ02,][]{Berry04}.  The roundness test is typically a quadrupole, with $m=\pm2$ azimuthal symmetry, in the sheared coordinate frame where the object is round.  Introduction of a covariance between $R$ and a hidden parameter with azimuthal order $m_H$ requires that the PSF have elements with azimuthal order $m_\star = \pm m_H \pm 2$.  ``Selection rules'' for covariances induced by finite sampling and data range are more complex.

We can tolerate biases in the measurements of individual galaxy ellipticities if we can be assured that they will go to zero when averaged over the full (isotropic) population of pre-lensing galaxy shapes.  For the shear derived from a full population to be biased due to truncation of some parameter $H$, we need
\begin{equation}
\left\langle H_{\rm true} {\rm Cov}(H,R) \right\rangle \ne 0.
\end{equation}
Both $H_{\rm true}$ and the covariance must be evaluated in the sheared coordinate system in which the galaxy is round. 

Consider the case when $H$ is the coefficient of a monopole basis function with high radial order, which is truncated from the fit to the data.  For example VB attempt to fit a deVaucouleurs-profile elliptical galaxy with a basis set consisting of one (or a few) Gaussian ellipsoids.  In the frame where these galaxies appear round, the galaxies have a monopole residual to the Gaussian fit.  If the PSF is circular in the observed frame, then it is elliptical in this sheared frame, with an ellipticity 90$^\circ$ out of phase with the original galaxy ellipticity.  The $m=2$ components of this (sheared) PSF will induce correlations between the amplitude $H$ of the truncated monopole basis function and the quadrupole component of the fit.  These induced biases will always be aligned with the galaxy ellipticity.  The result will be a multiplicative shear calibration error.  

Underfitting can also result from unmodelled galaxy terms beyond the monopole.  For example if galaxies tend to be disky, they have $m=\pm4$ components with a well-defined phase relation to the galaxy ellipticity.  These terms are coupled to the roundness test by the PSF, potentially causing systematic over- or under-estimation of the ellipticity.

A circular PSF acting on an isotropic population of galaxies has no preferred direction and hence cannot induce an additive shear error.  If, however, the PSF has anisotropic components, particularly at $m=2$, then underfitting can induce additive errors by correlating the quadrupole roundness test with unmodelled monopole components of the galaxies.

\subsection{Circular-model methods}
Several shear-measurement methods assign ellipticities to galaxies by fitting the observed data with models of perfectly elliptical galaxies.  We note that these are subject to the same underfitting biases.  These methods work as follows:
\begin{enumerate}
\item Select a circularly symmetric model galaxy function $\psi_o({\bf x})$.  The function may have additional parameters for size, etc., but this is secondary to our discussion.
\item For a trial centroid ${\bf x}_0$, shape $\eta$, and flux scaling $f$, produce the model observed image $\psi({\bf x})=f \psi_o[{\bf S}_\eta^{-1}({\bf x}-{\bf x_0})] \otimes T.$
\item Calculate the likelihood ({\it e.g.} $\chi^2$) of the observed data matching the sheared model.
\item Find the shear, centroid, and flux that maximize the likelihood of the data.  The galaxy is assigned an ellipticity equal to the model shear that maximizes the likelihood.  Some methods choose to estimate the likelihood-weighted mean values rather than the maximum-likelihood values, but for the high-$S/N$ cases we consider now, the difference is unimportant.
\end{enumerate}
\citet{Kuijken99} and \citet{VB} use a method in which the circular model is a sum of Gaussians; \citet{lensfit} have an exponential or deVaucouleurs profile circular model.

The method will clearly succeed without bias (for high $S/N$) when the galaxies are indeed sheared versions of the proffered models.  However for more general galaxies, these methods will be subject to the same biases as the truncated-basis-set methods described above.  Consider the set of five functions 
\begin{eqnarray}
\nonumber
\psi_o, & & \\
\psi_x & = & {\partial\psi_0 \over \partial x}, 
\psi_y = {\partial\psi_0 \over \partial y}, \nonumber \\
\psi_1 & = & {\partial\psi_0[{\bf S}_{\eta}^{-1}(\bf x)] \over \partial \eta_1}, \nonumber \\
\psi_2 & = & {\partial\psi_0[{\bf S}_{\eta}^{-1}(\bf x)] \over \partial \eta_2}, \nonumber \\
\end{eqnarray}
such that we can approximate
\begin{eqnarray}
f\psi_0[{\bf S}_\eta^{-1}({\bf x}-{\bf x}_0)]  & \approx & f_0\psi_0({\bf x})  \nonumber \\
 & & + (f_0x_0) \psi_x({\bf x})
+ (f_0y_0) \psi_y({\bf x}) \nonumber \\
 & & + (f_0\eta_1) \psi_1({\bf x})
+ (f_0\eta_2) \psi_2({\bf x}).
\label{circbasis}
\end{eqnarray}
for small departures from a centered, circular model.
The circular-model-fitting technique will consider $x_0=y_0=\eta_1=\eta_2=0$ to be the correct center and shape for this galaxy if the likelihood ${\cal L}$ of the fit of (\ref{circbasis}) to the observed data satisfies
\begin{equation}
\left.{\partial {\cal L} \over \partial x_0}\right|_{x_0=0} = 0
\end{equation}
and likewise for derivatives with respect to $y_0, \eta_1,$ and $\eta_2$.

However we can also view the five functions in (\ref{circbasis}) as the first elements of some complete basis set, with the coefficients of the last four terms serving as our centroid and roundness null tests.  The basis-set method will consider the galaxy to be round and centered in this frame if the likelihood of the fit of (\ref{circbasis}) to the data is maximized by nulling the coefficients of the last four terms.  This is exactly the same condition for shape assignment as in the circular-model-fitting technique.  Hence we can consider the latter technique to be a special case of fitting with a truncated basis set.

MBLB report $\approx20\%$ errors in measurement of deVaucouleurs-profile elliptical galaxies using Gauss-Laguerre measures of ellipticity---or worse with some other methods.  MBLB report that these large errors may be attributable to a degenerate behavior of the fitting method: with very well-resolved deVaucouleurs-profile galaxies, the condition used to defined the size scale of the Gauss-Laguerre basis set may lead to attempts to measure shapes by fitting the central intensity spike of the galaxy.  The basis functions can then be poorly resolved
by the image pixel scale.  We do not investigate in this paper the domain where the PSF-convolved image has structure well below the pixel scale, leaving this for future work.

\section{Potential solutions}
\begin{figure*}
\includegraphics[width=1.99\columnwidth]{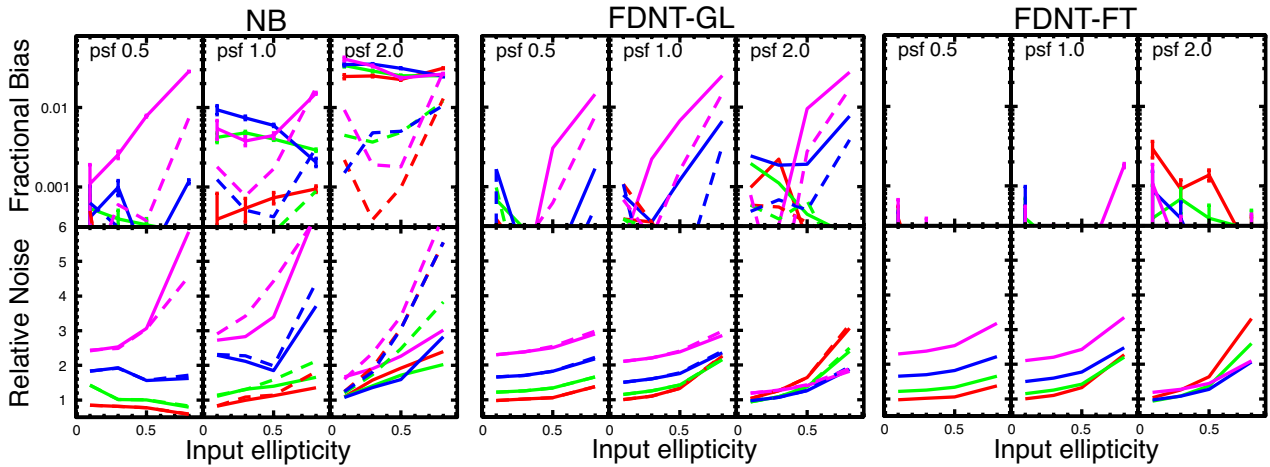}
\caption[]{Biases and noise levels for three shape measurement methodologies.  In each panel the upper row gives the fractional error $|\delta e| / e$ on the measured ellipticity of a Sersic-profile galaxy with input ellipticity $e$ labelled on the horizontal axis.  The curves are red, green, blue, and magenta for Sersic index $n=0.5, 1, 2,$ and 4.  The left three panels use the Gauss-Laguerre basis-set fitting method of NB07, deconvolving the apparent galaxy ellipticity from a Gaussian PSF with ellipticity $0.14$ and half-light radius of half, equal, or double that of the galaxy, as marked.  The deconvolved galaxy shape is modelled with order $N=8$ (solid) or $N=12$ (dashed) decomposition.  The middle set uses the FDNT shape measurement method introduced in this paper, executing the Fourier transform via a ``native'' GL decomposition of order 12 (solid) or $N=16$ (dashed).  The right-hand set of panels shows biases using the FDNT method with FT of the real-space data.  The lower panel shows the noise level in the inferred ellipticity for each case, relative to ellipticity error for a circular Gaussian galaxy with the same half-light radius and background noise as each test case, and therefore indicate how each measurement is being degraded by the ellipticity and non-Gaussian galaxies.  We see that the FDNT methods dramatically reduce the shape measurement bias and with noise levels that are lower and less affected by order of the expansion.  Note many FDNT measurements have errors below the scale of the plot, of the order $10^{-4}$, at the limit of precision of these tests.
}
\label{biases}
\end{figure*}

The left panel of Figure~\ref{biases} plots the bias in $e$ measurements of Sersic-profile galaxies viewed at very high $S/N\approx10^4$ with a Gaussian PSF, as measured with the Gauss-Laguerre (GL) basis function fitting method described in \citet[NB07]{NB07}.  We see that galaxies with low Sersic indices, close to the Gaussian value of $n=0.5$, are measured with shape biases below the 1 part in $10^3$ threshold, as long as they are well resolved.  However biases of up to several percent appear when the Sersic index rises toward the deVaucouleurs value of $n=4$, when the galaxies become highly elliptical, and when the galaxies are more poorly resolved.   All three of these factors lead the galaxy to be more poorly represented by the finite GL basis set, opening the door to underfitting biases.

One way to avoid the underfitting bias would be to find a basis set in which all galaxies are compact and represented without significant bias by a finite number of coefficients, {\it e.g.}
\citet{Ngan} propose a basis set generated from Sersic profiles, which may represent Hubble-sequence galaxies more compactly than the Gauss-Laguerre basis sets used by NB07 and \citet{RB03}.  A viable methodology from a Sersic basis has not yet been reported, however.  The less regular morphology of fainter galaxies may make it difficult to obtain a sufficiently general basis.  We will not pursue this approach in this paper.

A second strategy is to simply extend the basis set to higher order until the biases are reduced to desired level.  The dashed lines in the left panel of Figure~\ref{biases} show the impact of increasing the GL expansion from order $N=8$ (45 basis functions) to $N=12$ (91 basis functions).  The biases are indeed substantially reduced in most cases, but remain unacceptably high for poorly resolved objects and for objects with higher $n$ and high ellipticity.  We note however that a price has been paid, particularly for the poorly resolved objects, in that the resultant uncertainties in the ellipticity have risen substantially (see lower row of plots).  While this is not a problem for an object with $S/N=10^4$, it will be a problem for real observations, where $S/N$ is expensive to obtain. 

Why does going to higher order lead to higher noise primarily in the poorer-resolution cases? Recall that in the basis-set-fitting method, four basis functions are designated as the centroid and roundness tests.  Considering just one of these coefficients $b_i$ and its associated function $\psi_i$, we know that as the basis set becomes more complete, the least-squares solution will converge (at high $S/N$) to the desired limit
\begin{equation}
b_i \rightarrow \int d^2k\, \tilde\psi_i({\bf k}) \tilde I_a({\bf k}).
\end{equation}
We express the integral in Fourier space because it is easier to understand the noise properties of convolved images in Fourier space.  We also work in a coordinate system where the object is nearly round, so that we are testing for $\eta=0$ and we can for the moment omit the ${\bf S}_\eta$ matrices for clarity.
For background-limited images, the {\em observed} image $I_o({\bf k})$ has a uniform (white) noise spectrum $N$.  However to retrieve the pre-seeing image $I_a$, we must divide by the modulation transfer function (MTF), which is the Fourier conjugate of the PSF: $\tilde I_a({\bf k}) = \tilde I_o({\bf k}) / \tilde T({\bf k})$.  This means that the effective noise power on the pre-seeing image is $N / |\tilde T^2({\bf k})|$, rising to infinity at higher $k$.  We thus expect that as the basis becomes complete, the measurement of the roundness test $b_i$ will acquire noise level
\begin{equation}
{\rm Var}(b_i) = N \int d^2k\, \left | \tilde\psi_i({\bf k}) / \tilde T({\bf k}) \right|^2.
\label{varb}
\end{equation}
If we have chosen a roundness test $\tilde\psi_i({\bf k})$ which goes to zero before the inverse MTF $1/\tilde T({\bf k})$ blows up, then the noise on $b_i$ will remain bounded as we increase the fit to high order.   We can reduce bias by fitting to higher order without paying a stiff noise penalty
only if we have chosen our roundness test to roll off more quickly than the MTF in $k$-space.

\begin{figure}
\includegraphics[width=0.99\columnwidth]{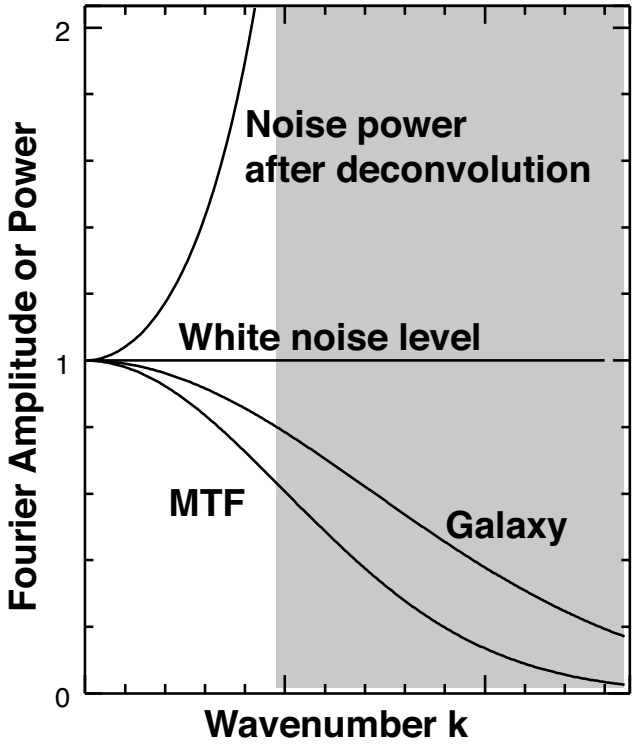}
\caption[]{Illustration of the difficulty of model fitting to a poorly resolved galaxy.  If a galaxy is smaller than the PSF, then as shown here its Fourier transform extends past the MTF, which is the Fourier transform of the PSF.  If we try to measure the pre-seeing shape of the image, we must divide the observed image by the MTF, which raises the power spectrum by the inverse square of the MTF, noted by the upper line.  In the gray region of $k$ space the image is very noisy.  If we are testing galaxy roundness with a filter that is matched to the galaxy size, we will obtain a very noisy answer because the filter extends into the degraded zone.
}
\label{kbias}
\end{figure}

If, however, we are using a roundness test that extends to regions where $1/\tilde T$ is large, then we will find the error on $b_i$ climbing higher and higher as we use more complete basis sets.  As illustrated in Figure~\ref{kbias}, this is precisely what happens when using a model-fitting technique on an unresolved galaxy.  One would like to choose a basis function $\psi_i$ which is of similar size to the (pre-seeing) galaxy $I_a$ so that the representation of this galaxy is compact in the basis.  However if the galaxy is smaller than the PSF, then in Fourier domain, the roundness test $\tilde \psi_i$ will extend to higher $k$ than the MTF $\tilde T$, and the roundness test noise will diverge as the basis set becomes complete.

When we use a roundness test that extends to $k$ values where the $1/\tilde T$ deconvolution term becomes large, we face a tradeoff between noise and bias: as we truncate the basis set to avoid fitting degeneracies at high $k$, we become subject to bias if the truncated basis does not properly extrapolate behavior to high $k$.  We conclude that underfitting bias stems, at root, from an attempt to measure galaxy shapes using information at high $k$ which is not present in the data.

The MTF is not the only reason why our knowledge of a galaxy's $I_a$ is incomplete.  Finite sampling can cause aliasing ambiguities.  In real space, missing pixels, and the necessarily finite measurement aperture for each galaxy, are also forms of missing information.  We rely upon a finite-basis representation of the galaxy to fill in the missing information during a shape measurement.  In the absence of a known finite basis representation of the galaxy population, our best strategy is to select a roundness test that is least sensitive to poorly observed regions of $x$ or $k$ space.

\section{Fourier-domain null test method}
\subsection{Designing the best roundness tests}
We propose a new shape-measurement method, ``Fourier-domain null testing'' (FDNT), which explicitly chooses a roundness test that does not use information destroyed by the MTF.  We work in the Fourier domain where the effect of the PSF is more transparent. We assume that the roundness test is linear and a pure quadrupole.  Working in the coordinate frame where we are testing for $\eta=0$ the tests are:
\begin{eqnarray}
t_1(I_a) & \equiv & \int d^2k\, \tilde I_a({\bf k}) \tilde w(k) (k_x^2-k_y^2) \\
  & = & \int d^2k\, \left[\tilde I_o({\bf k}) / \tilde T({\bf k})\right]  \tilde w(k) (k_x^2-k_y^2) \\ \nonumber
t_2(I_a) & \equiv & \int d^2k\, \tilde I_a({\bf k}) \tilde w(k) (2k_xk_y). \\ \nonumber
\end{eqnarray}
Equation~(\ref{varb}) can be used to estimate the variance of $t_i$ in the case of white noise in the image.  The response of the roundness test to a small shear is our desired signal; BJ02 show that when the galaxy is nearly round, this will be related to the unlensed intensity $I_u$ by
\begin{equation}
{dt_1 \over d\eta_1} \propto \int d^2k\, k^2\tilde w(k)
k { d \over dk} \left\langle \tilde I_u({\bf k}) \right\rangle_\theta
\end{equation}
The brackets indicate averaging around the azimuth angle.
Combining the expressions for signal and noise of $t_i$ we can find the radial weight $\tilde w(k)$ that produces the Fourier-domain quadrupole roundness test with the best $S/N$ on shear:
\begin{eqnarray}
\tilde w(k) & = & \tilde w_g(k) \tilde T_0^2(k), \\
\label{T0def}
\tilde T_0^2(k) & \equiv & \left\langle \left| \tilde T({\bf k})\right|^{-2} \right\rangle_\theta^{-2} \\
k^2 \tilde w_g(k) & \equiv & k { d \over dk} \left\langle \tilde I_u({\bf k}) \right\rangle_\theta
\end{eqnarray}
The $\tilde w_g$ term in the weight function serves to weight those annuli in $k$ space where we expect the galaxy to exhibit the largest signal under shear.  For a Gaussian galaxy with $I_a \propto e^{-r^2/2\sigma_g^2}$, the roundness-test weight should also be a Gaussian, $\tilde w_g(k) = e^{-k^2\sigma_g^2/2}$.  BJ02 show that this choice is also near optimal for an exponential-disk galaxy (Sersic $n=1$).  We will retain the Gaussian form of $\tilde w_g$ even though it is suboptimal for steeper-cored Sersic profiles, because the details are not critical and any well-behaved function yields a valid roundness test.

More critical is the $\tilde T_0^2(k)$ term in the weight, which serves to roll off the weight wherever the MTF is getting small.  Note that the roundness test goes to zero at any radius where the MTF drops to zero, so our roundness test never requires extrapolation into unmeasured regimes.  Assuming that we have access to a Fourier representation of our observed galaxy, we {\em do not need to fit the galaxy to any basis set} to execute the roundness test.  The tests for centroiding and roundness at shear $\eta$ are
\begin{eqnarray}
t_j & = & \int d^2k\, \left[\tilde w_j(|{\bf S_\eta k}|) / \tilde T({\bf k})\right] \tilde I_o({\bf k}) \nonumber \\
\tilde w_j(k) & = & \tilde w_g(k) \tilde T_0^2(k) \times
\left\{
\begin{array}{cl}
k_x^2 - k_y^2 & \quad j=1 \\
2k_x k_y          & \quad j=2 \\
i k_x                & \quad j=x \\
i k_y                & \quad j=y
\end{array}
\right.
\label{test1}
\end{eqnarray}
To assign an ellipticity to the galaxy, we first transform the {\em observed} galaxy $I_o({\bf x})$ into Fourier domain, as well as the local PSF.  For a given trial shear $(\eta_1, \eta_2)$ we can readily create the four bracketed functions in (\ref{test1}), to evaluate the four null tests.
Then we iterate the coordinate origin and the shear components $(\eta_1, \eta_2)$ until all the tests are zero.  We assign the galaxy an ellipticity $e=\tanh \eta$ at the null.

The right-hand panel of Figure~\ref{biases} shows the results of applying FDNT shape measurements to the same perfectly elliptical Sersic-profile galaxies measured with the GL basis-set technique.  There is dramatic improvement: in nearly all cases the biases are too low to even appear on our plots.

\subsection{MTF shrink factor}
\begin{figure}
\includegraphics[width=0.99\columnwidth]{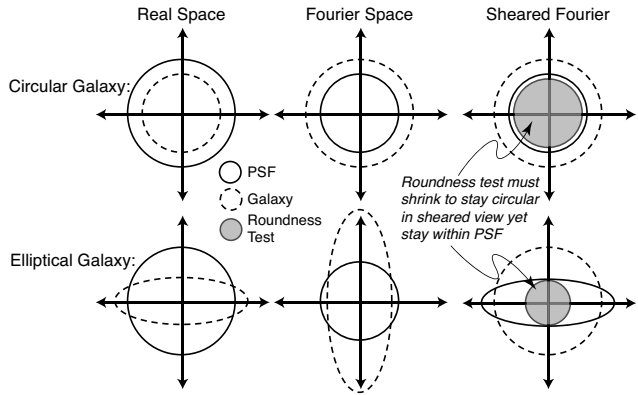}
\caption[]{Illustration of PSF ``shrinkage'' in $k$ space for elliptical galaxies.  We schematically draw the size of a galaxy and PSF in real space, in the Fourier domain, and in Fourier domain after shearing the coordinates to make the galaxy appear round.  It is in this last frame that we want to apply a circular ``roundness test'' weight function in order to have well-behaved shape assignments.  In the bottom row we see how a more elliptical galaxy forces the shaded roundness test to be smaller (in $k$ space) if it is to remain inside the MTF where the data are not degraded.
}
\label{psfshrink}
\end{figure}

Note in the lower panels of Figure~\ref{biases} that more elliptical galaxies have higher shape noise levels (expressed as errors in $\eta$), {\it i.e.} are less sensitive to applied shear. There is a subtle reason for this.  The FDNT weight function contains a circularized MTF term $\tilde T_0^2(k)$ defined in \eqq{T0def}.  This averaging must be done in the sheared coordinate system where we are testing for roundness.  Even if the PSF is circular on the observed image, it will not be circular in the sheared coordinates.  Furthermore the function $\tilde T_0^2$ as defined in \eqq{T0def} will change subtly as we test for roundness at different $\eta$ values.  In particular,
if the MTF goes to zero at some $k_{\rm max}$ in the observed frame, then the function $\tilde T_0^2$ should go to zero at a smaller radius $k_{\rm max}/s$, where $s$ is a ``shrink factor'' given by $s=\exp(\eta/2) = [(1+e)/(1-e)]^{1/4}$.  As illustrated in Figure~\ref{psfshrink}, shearing the coordinate system makes the PSF broader in one dimension, effectively degrading resolution in this direction.  But since our roundness test requires a circularly symmetric weight function, we must shrink the entire weight function to fit it into the useful region of the now-elliptical MTF.  Because there is now less $k$ space to use for the roundness tests, the shape measurement becomes noisier.

Rather than recompute the circularized $\tilde T_0^2$ via (\ref{T0def}) at every shear being tested, we simplify by computing and tabulating a function ${\cal T}(k)$ defined to be $\tilde T_0^2$ in the observed coordinate frame ($\eta=0$).  Then for non-zero $\eta$ we just calculate the shrink factor $s$ and use the function ${\cal T}(sk)$ for the PSF part of the weight.

A second effect of the distortion of the MTF as we change $\eta$ is that the roundness tests are {\em not} invariant under shear of the underlying galaxy.  A galaxy that is sheared to become more elliptical will be nulled by a roundness test that is slightly smaller (in $k$ space) than the unsheared galaxy.  This violates one of our original criteria for roundness tests and leads to potential multiplicative shear biases when the source galaxies have radial ellipticity gradients, a problem which we elaborate and solve in \S\ref{eg}.  These problems are not specific to the FDNT method; all practical shape measurements apply some kind of PSF-dependent weighting to galaxies, and this weight will vary under shear unless we specifically design the weighting scheme to avoid this.  

\subsection{Implementation notes}
The FDNT measurement can be quite fast, requiring an FFT of the pixelized $I_a$ into Fourier domain, and the evaluation of the four weight functions at each point in the transformed image.  The derivatives of the null tests with respect to centroid position are easily calculated; we solve for the centroid position by assuming linear response to centroid shifts relative to the starting position.  The search for the shear that then nulls $t_1$ and $t_2$ can use standard nonlinear root-finding techniques, in our case Powell's ``dogleg'' method \citep{MNT}.

If the observed image has stationary noise, then the covariance matrix of the Fourier components is diagonal, and it is also very fast to propagate the image noise into a covariance matrix for the 4 null tests.  It is however computationally expensive to propagate a non-uniform noise map into Fourier domain, and therefore the FT method is ill-suited to estimating uncertainties on the null tests in this case.

\subsection{Fourier transform via basis sets}

Other shape-measurement methods make use of Fourier-domain representations of galaxy images \citep{ZhangFourier,lensfit,HB09}.  It may prove difficult, however, to turn such methods into practical weak-lensing algorithms, because the Fourier transform is not robust to missing pixels, {\it e.g.} from cosmic rays.  Furthermore, galaxies have neighbors, so we have a limited region over which to perform a transform without running into neighbor galaxies, which places a limit on the $k$-space resolution one can obtain.   Zero-padding a small postage stamp can increase the resolution of the FT, but one is then assuming that the galaxy flux drops identically to zero outside the postage stamp, which may induce a bias in the shape measurement.

We need some way to interpolate in real space to missing pixels or extrapolate to large radii.  We can call upon fitting methods again to do this.  If we fit the {\em observed} image to a finite basis set such that $I_o({\bf x}) = \sum b_i \psi_i({\bf x})$, then in Fourier domain the galaxy is
$\tilde I_o({\bf k}) = \sum b_i \tilde\psi_i({\bf k})$.  The roundness (or centroid) test $t_j$ is
\begin{eqnarray}
t_j & = & \sum_i \beta_{ji} b_i, \\
\beta_{ji} & \equiv &
\int d^2k\, \left[\tilde w_j(|{\bf S_\eta k}|) / \tilde T({\bf k})\right] \tilde \psi_i({\bf k}), \\
{\rm Cov}(t_j, t_k) & = & \sum_{lm} \beta_{jl} {\rm Cov}(b_l, b_m) \beta_{mk}.
\end{eqnarray}
Thus we need fit the observed image $I_o({\bf x})$ just once to obtain the $b_i$ and their covariance matrix, {\it i.e.} we project the pixel data onto a description $b_i$ and execute the tests on the $b_i$.  If $b_i$ is of lower dimension than the pixel count, we have made the propagation of errors easier than in the FT method, and we become robust to missing pixels.

We have implemented a version of the FDNT tests which uses the elliptical GL functions as a basis set.  These basis functions have the advantage of being their own Fourier transforms (up to a power of $i$), which makes the GL basis fit a fast way of bringing the galaxy into the Fourier domain.  The fit is robust to missing pixels if the order of the GL decomposition is not too high, and the GL functions are compact in real (and Fourier) space so they do not need very large postage stamps to fit successfully.

The central panel of Figure~\ref{biases} shows the result of executing the FDNT method with Fourier transforms done by the GL decomposition.   We see that for Sersic indices $n=2$ and $n=4$, the GL transform induces biases that are worse than our FDNT-FT method and exceed the desired $10^{-3}$ threshold when the source ellipticity gets high.  These galaxies are of course the ones that are least compactly represented by the elliptical GL basis, and hence the execution of the Fourier transform via GL decomposition leads to biases.  We would find larger biases if the PSF were an Airy function instead of a Gaussian, because it is known that the GL decomposition requires very high order to reproduce the $k=0$ cusp of the Airy MTF (NB07).   We have once again the situation that the use of a finite basis set to fill in missing galaxy information can lead to biases if the basis set is not well matched to the galaxies.

However we should note that the performance of the FDNT-GL method, while worse than FDNT-FT, is much better than the performance of the NB07 implementation of the basis-set fitting method.  Note that the biases are much smaller than for the NB07 method, and also the shear noise level remains stable when we increase the order of the GL decomposition from 12 to 16, even for poorly resolved galaxies, and the noise levels are generally lower.  The FDNT-GL method differs from basis-set-fitting in that we are fitting the {\em observed} object profile, not attempting to fit the {\em pre-seeing} image.  This makes the fit more stable so we can go to higher order.  We can also use a basis that is matched in size and shape to the object, making the galaxy more compact in the basis.  And by using the FDNT roundness test instead of one of the basis coefficients, we make our shape measurement immune to noise that comes from attempting to fit the pre-seeing galaxy at $k$ values suppressed by the MTF.

We are not constrained to using the GL basis set; one can execute the Fourier transform by fitting any desired basis set to the observed image.  The direct FT method has the advantage of lower biases (and faster execution) than the GL FDNT method.  However the benefit of projecting the pixel data onto a basis set before transforming is that we can compress the pixel data to a smaller number of parameters, making propagation of errors more tractable, and also making the measurement robust to missing pixels or smaller measurement apertures.

\section{Ellipticity gradients}
\label{eg}
VB find that a galaxy composed of a superposition of bulge and disk components with different ellipticities was measured with 2\% multiplicative shear error, and that this error was not reduced by increasing the radial freedom of their circular-model-fitting method (see their Figure 7).  This multiplicative error is a consequence of two circumstances.  First, the galaxy's measured ellipticity varies with the scale of the weight function applied to the calculation of its quadratic moments, {\it i.e.} its core is rounder than its outskirts, or vice-versa in Fourier space.
Second, the effective weight function of the circular-model-fitting method changes implicitly with the ellipticity $\eta$ being applied to the model as the MTF gets sheared in the frame where the model is circular.  As discussed above, this violates the prescription that our roundness test should be invariant under applied shear.  We reiterate that this implicit change in the shape weighting function under shear is present in nearly every shape-measurement methodology published to date, with the choice of weight scale often buried deep in the implementation details.    The resultant biases are probably endemic to current shape measurements.  Here we investigate the problem and its solution in more detail within the context of the FDNT method.

\subsection{Estimation of bias}
\label{egestimation}
Consider an unlensed, pre-seeing galaxy that when measured with a null test of scale $s$ yields an ellipticity $(\eta_1\ge0, \eta_2=0)$.  If the galaxy is not perfectly elliptical, it may have nonzero derivatives $d\eta_1 / ds$ and $d\eta_2/ds$ with respect to a change in the weight scale factor.  We will say that such a galaxy has ``ellipticity gradients'' even though its isophotes need not be ellipses.

Assume that the weight scale $s$ applied during a null test depends upon the amplitude $\eta_t$ of the shear at which we're looking for the null.
In most function-fitting methods, this change in scale size is implicit and not controlled, whereas we will control this explicitly in the FDNT method.
There is some potentially non-zero derivative $ds/d\eta_t$ in most shape-measurement methods.

Now assume that the shrink factor was $s(\eta_1)$ when the galaxy was unlensed, and then we apply a small lensing shear $(\Delta \eta_1, \Delta \eta_2)$ to the galaxy and we remeasure it.  The ellipticity $(\eta^\prime_1, \eta^\prime_2)$ measured by the method will satisfy
\begin{eqnarray}
\label{ep1}
\eta^\prime_1 & = & \left( \eta_1 + {d\eta_1 \over ds} \Delta s \right) + \Delta\eta_1 \\
\label{ep2}
\eta^\prime_2 & = & \left(  {d\eta_2 \over ds} \Delta s \right) +{\eta_1 \over \tanh \eta_1} \Delta\eta_2 \\
\Delta s & \equiv & s(\eta^\prime) - s (\eta_1) \approx {ds \over d\eta_t} (\eta^\prime_1 - \eta_1)
\end{eqnarray}
The key observation is that the parenthesized quantities in Equations~(\ref{ep1}) and (\ref{ep2}) are the shape of the {\em unlensed} galaxy if measured with the slightly different scale $s+\Delta s$ that will be used when measuring the {\em lensed} galaxy.  The right-most terms are the differential to the ellipticity caused by the applied lensing shear, as per Equation~(\ref{etransform}).\footnote{Note that we are describing the galaxy ellipticities with $\eta^\prime$ instead of the $e^\prime=\tanh\eta^\prime$ used in Equation~(\ref{etransform}).}  The applied shear shifts the measured shear according to
\begin{eqnarray}
\label{egeffect1}
\delta \eta_1 \equiv \eta^\prime_1 - \eta_1 & = & \Delta\eta_1 \left[ 1 - {d\eta_1 \over ds} {ds \over d\eta_t}\right]^{-1} \\
\delta \eta_2 & = & {\eta_1 \over \tanh \eta_1}\Delta\eta_2 + {d\eta_2 \over ds} {ds \over d\eta_t} \delta\eta_1.
\end{eqnarray}
The ellipticity gradient hence causes the measured response to shear to depart slightly from the applied shear.  Because the coordinate system was chosen to align with the galaxy ($\eta_2=0$), a change in $\eta_2$ is a rotation of the ellipticity---we would
find the change in position angle $d\beta/ds = (2 / \eta_1) d\eta_2/ds.$  This is a quantity that changes sign under reflection of the unlensed galaxy about the $x$ axis.  Since the unlensed population of galaxies should be parity-invariant, we can assume that $\langle d\eta_2/ ds \rangle=0$ when averaged over the population, so the ellipticity-gradient effect on $\delta\eta_2$ averages to zero. 

The ellipticity gradient therefore causes a rescaling of the measured shear $\delta \eta_\parallel$ component lying parallel to the galaxy's major axis, while leaving the orthogonal shear component $\delta\eta_\perp$ unchanged on average.  To gauge the effect on the weak lensing shear measurement, we average the measured differential shear over a population of galaxies with unlensed ellipticity $\eta$ and random position angles $\beta$.  After some algebra to execute the average over $\beta$ we obtain, for both shear components $j\in \{1,2\}$,
\begin{eqnarray}
\left\langle  \delta \eta_j \right\rangle_\beta & = & {\cal R}_1 \Delta \eta_j  \\
\label{r1def}
{\cal R}_1 & = & {1 \over 2} \left[ {1 \over 1 - EG}
    + {\eta \over \tanh \eta} \right] \\
EG & \equiv & {d\eta_\parallel \over ds} {ds \over d\eta_t}.
\label{egfix}
\end{eqnarray}

\subsection{Correction for ellipticity gradients in FDNT}
In the FDNT method, the weight factor is constructed explicitly from two $k$-domain factors, the galaxy function $\tilde w_g(k)$ and the circularized PSF factor $\tilde T_0^2(k)$.  If either changes when the galaxy is sheared, we will incur an ellipticity gradient bias.  We have already described that we replace the $\tilde T_0^2$ term with ${\cal T}(ks)$ with $s=\exp(\eta_t/2)$ to account for shrinkage of the MTF minor axis in the circularizing coordinate frame.  We therefore have
$ds/d\eta_t= \exp(\eta_t/2) / 2$.

We also need to control any change in $\tilde w_g(k)$ under shear of the galaxy.  We first set $\tilde w_g(k) = \tilde W_g(k\sigma_g)$ where $\tilde W_g$ is a fixed function form---a Gaussian---with explicit scale $\sigma_g$.  We choose $\sigma_g$ by requiring solution of this equation:
\begin{eqnarray}
\int d^2k\, k^2 {\cal T}(ks) \tilde W_g(k\sigma_g) = & & \nonumber \\
\qquad 
\int d^2k\, k^2 {\cal T}(ks) \left[ \tilde I_o({\bf S}_\eta^{-1} {\bf k}) / \tilde T({\bf S}_\eta^{-1} {\bf k}) \right]. & &
\label{sigmag}
\end{eqnarray}
This condition requires that, in the sheared coordinate system which make the galaxy round, the $k^2$ moment of the weight function must match that of the galaxy itself.  Both functions are weighted by the factor ${\cal T}(ks)$ in order to confine the calculation to the domain where the measurement survives convolution with the PSF.  Because the condition is applied in the circularizing coordinate frame, the chosen $\sigma_g$ should be invariant under applied shear at fixed $s$.  

The roundness-test weight  is therefore scaled by $s$ both directly through ${\cal T}(ks)$ and indirectly through any potential variation $d\sigma_g/ds$ implicit in \eqq{sigmag}.  We can calculate the resultant $d\eta_\parallel / ds$ either by propagating derivatives of the roundness tests with respect to $s$ and $\sigma_g$, or by intentionally perturbing $s$ and repeating the solution for the $\eta$ which nulls the tests.  In particular, if ${\bf t}$ is the vector of null tests, then the measured ellipticity ${\bf e}$ of the galaxy varies with $s$ as
\begin{equation}
{d\bfe \over ds}  =  - \left({d{\bf t} \over d\bfe}\right)^{-1} {d{\bf t} \over ds}.
\end{equation}
We can then take the component of this derivative in the direction parallel to the galaxy ellipticity to get $d\eta_\parallel/ds$ and derive the ellipticity-gradient response factor $EG$.

\subsection{Testing}
For galaxies that are not perfectly elliptical, there is no single correct measure of the galaxy ellipticity, and we must perform a ``ring test'' as described by NB07 in order to test for shear measurement biases.  Copies of the unlensed galaxy are drawn with rotation $\theta$ uniformly distributed from 0--2$\pi$.  Each copy is then sheared by an amount $\Delta\eta$, taken here to be along the $x$ axis.  The copies are then convolved with the PSF, sampled on the pixel grid, and given appropriate noise levels.  Each copy is then measured using the FDNT-FT method to yield galaxy shape components $\eta_{1,i}$ and $\eta_{2,i}$ and an estimate of the ellipticity-gradient factor $EG$.  From $EG$ and the measured shape $\eta$ of each galaxy we also obtain an estimate ${\cal R}_{1,i}$ of the shear responsivity in \eqq{r1def}.  From the ring ensemble we then estimate the applied shear as
\begin{equation}
\widehat{\Delta\eta} = {\sum \eta_{1,i}
\over 
\sum {\cal R}_{1,i}}
\label{eest}
\end{equation}
If this estimator reproduces the shear $\Delta\eta$ applied to the ring without bias, then it will also produce an unbiased shear estimate from any source population that is isotropic before lensing. 
When the ring test is executed with applied shear $\Delta\eta=0$, we always find the estimator consistent with zero, indicating that we have no additive biases arising from the non-circular PSF.
We search for multiplicative bias by seeing how well an applied shear of $\Delta\eta=0.02$ is retrieved. 

We do not report the mean of the estimated component $\eta_2$ orthogonal to the applied shear.  Symmetry considerations imply that the mean $\eta_2$ will vanish when the source population is invariant under reflection about the $x$ axis---and this does occur with ring test for Test Galaxies 1 and 2.  Test Galaxy 3 does not have the reflection symmetry, so ellipticity gradients or other biases could cause $\langle \eta_2 \rangle \ne 0$.  However
any Universe containing galaxies like this would contain an equal number of mirror-imaged versions of Test Galaxy 3.  As argued in \S\ref{egestimation}, this will cancel $\eta_2$ biases from ellipticity gradients.

We note that the ``perfect'' ellipticity transformation \eqq{etransform} and the resultant responsivity estimate (\ref{r1def}) take a linearized response of the galaxy shape to an applied shear.  For any measure of shear, there are also terms cubic in the shear which are significant if we have applied shear $\Delta\eta\approx0.05$ and wish to measure $\Delta\eta$ to $<1$ part in $10^3$.  In the Appendix we give the third-order responsivity ${\cal R}_3$ that must be included to measure shears below the part-per-thousand level.  The results presented in this section apply the third-order corrections expected for a perfect shape measurement method to the ring test results.  The cubic terms derived in the Appendix also include transformation of the estimated shear from the unit-determinant $\Delta\eta$ to the traceless $\gamma$ measure.

\begin{figure}
\includegraphics[width=0.99\columnwidth]{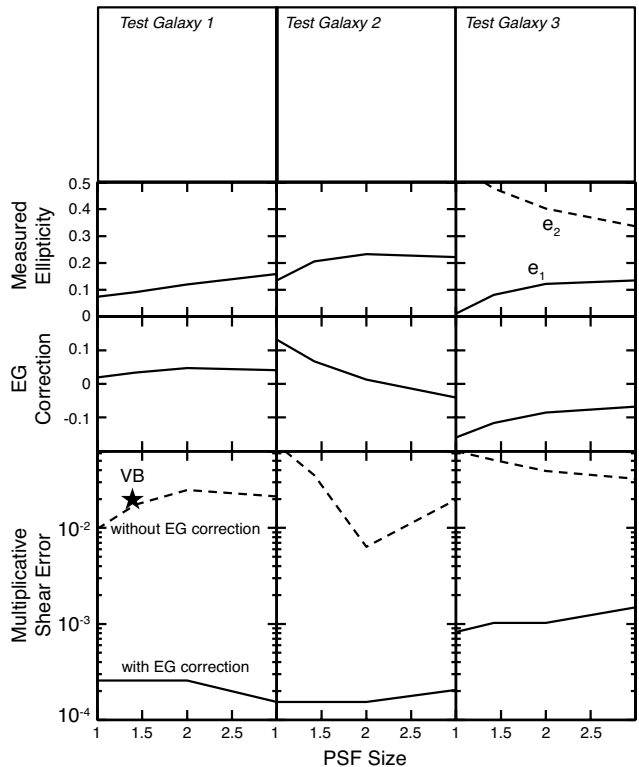}
\caption[]{ Multiplicative errors in the shape measurement of a simulated galaxies, having an ellipticity gradient, using the ``FDNT-FT'' method described in the text.  Each test galaxy has an outer exponential disk and an inner deVaucouleurs-profile ``bulge,'' with the two components having different ellipticities and scale lengths so that the overall ellipticity of the galaxy depends on the scale of the measurement weight.  For each test galaxy, the second row plots the ellipticity measured vs half-light radius of the PSF.  The third row plots the factor $EG$, measured from the data, which should be applied to the FDNT-FT shapes to infer the applied shear as per \eqq{r1def}.  The bottom row plots the multiplicative error on the inferred lensing shear both with and without the application of the $EG$ correction.  
Test Galaxy 1 is the ``spiral'' galaxy from \citet{VB}, with a mild change in axis ratio, and the multiplicative errors are negligible after $EG$ correction.
Test Galaxy 2 is a harsher test, with more elliptical and orthogonal components, and still has negligible error after correction.  Test Galaxy 3 is a yet more difficult case, with multiple maxima, a larger ellipticity gradient, and no inversion symmetry---even in this case, the $EG$ correction improves the multiplicative error 20-fold, to $\approx 10^{-3}$.
}
\label{egfig}
\end{figure}

We first perform a ring test using the FDNT-FT method on the simulated spiral galaxy of VB.  This galaxy has 20\% of its flux in a ``bulge'' with Sersic index $n=1.5$ and ellipticity $e=0.05$ and half-light radius of 1.5~pixels.  The ``disk'' component has $n=1$, $e=0.2$, and half-light radius 11.8 pixels, with the same position angle as the bulge component.  This galaxy is drawn in the top left panel of Figure~\ref{egfig} and its measured ellipticity is plotted in the panel below.  The measured ellipticity changes as we alter the size of the applied PSF and hence the scale of the null test weight function.  The dashed line in the lower panel shows the multiplicative error on the applied shear as measured by the ring test with no correction for the ellipticity gradient.  The star marks the 2\% bias of the VB method, slightly more biased than our uncorrected FDNT method.  If, however, we calculate the ellipticity gradient correction $EG$ (third panel) and apply it to the responsivity as in \eqq{r1def}, the error in the inferred shear is reduced 100-fold to negligible levels of $<3\times10^{-4}$, as recorded in the bottom solid line.

We conduct tougher tests by producing galaxies with larger ellipticity gradient corrections.
In the middle panel of Figure~\ref{egfig} the two Sersic components have orthogonal ellipticities and the required $EG$ correction is up to $4\times$ larger than for the VB galaxy, yet the $EG$-corrected shear estimate is still accurate to $<3\times10^{-4}$.  Test Galaxy 3 in the right panel
is a true torture test: the two components are not orthogonal and not concentric, so the galaxy has multiple maxima has no inversion or reflection symmetries, and has isophotal twists.  Without $EG$ correction, the multiplicative errors on shear are as large as 6\%, but the calculated $EG$ reduces bias at least 20-fold, close to the $10^{-3}$ target.

We speculate that the residual multiplicative errors in the torture-test are due to the limitations of linearized analysis of the ellipticity gradient effects in deriving \eqq{egeffect1}.  For the VB galaxy, $EG\approx0.03$, while $EG\sim0.1$ for the torture-test galaxy.  We should not be surprised to find shear errors at $O(EG^3\gamma)$ or $O(EG\,\gamma^3)$.  The the residual errors appear to indeed scale with applied shear as $\gamma^3$, but further investigation is required to check dependence on $EG$ and other galaxy or PSF properties.

We note that ``stacking'' methods will not suffer from ellipticity gradients, since stacking creates a mean galaxy image that tends toward circular at all isophotes \citep{Kuijken99,LewisStack}. A stack of $N$ galaxies with random (unlensed) orientations has ellipticity $\sqrt{N}$ smaller than the typical individual galaxy.  Hence the ellipticity gradient of the stack is reduced by $\sqrt{N}$ as well, as is the induced $EG$ bias.  For $N\ga 1000$, stacking will reduce the ellipticity-gradient bias as well as the $EG$ correction presented herein.  The relative success of stacking methods in the GREAT08 tests is likely a reflection of this advantage over other previous methods.
The FDNT method is the first proposal to properly correct for the ellipticity-gradient bias without stacking.

\subsection{GREAT08 tests}
We apply the FDNT-FT measurement method to test images comprising the low-noise sections of the GREAT08 Challenge. As extensively described in \citet{GR08Handbook} and \citet{GR08}, this challenge was to estimate the lensing shear applied to each of 15 images, with $10^4$ simulated galaxies of $S/N\approx 200$ in each image.\footnote{See {\tt http://www.greatchallenges.info} for documentation and images from GREAT08 and its successors.}  GREAT08 results are scored by $Q\equiv 10^{-4}/\sigma^2_{\gamma, \rm RMS}$, where $\sigma_{\gamma, \rm RMS}$ is the mean error in the estimate of each component of shear.

We found that the postage stamps on which the GREAT08 galaxies were drawn were small enough to truncate many of the more elliptical test galaxies, and also too small to contain the real-space versions of the filters being selected by the FDNT-FT algorithms.  D. Kirk and S. Balan kindly produced a new realization of the low-noise blind GREAT08 test suite, drawn from the same galaxy distributions as the original, but placed onto larger postage stamps.  The FDNT-FT algorithm was not provided with any knowledge of the characteristics of the underlying galaxy population, and we were blind to the input shears for the ``large-stamp'' test images.  We did, however, take advantage of the exact analytic form provided for the PSF.

The best performance reported for any method in \citet{GR08} was $Q\approx500$. \citet{Gruen} have subsequently reported $Q\approx1500$ for some subsets of the GREAT08 images, using a neural network to calibrate shear measurements to simulations of known input shear.
Without any training or recalibration, the FDNT-FT measurements achieved $Q=2997$, corresponding to an RMS shear estimation error of $0.0018$, which is only 20\% higher than the expected statistical uncertainty due to noise in the input images.  This $Q$ value is $6\times$ higher than any other reported in \citet{GR08}, and $3\times$ higher than the value estimated for the shear accuracy required to have negligible degradation of the statistical errors of hemisphere-scale WL surveys.

\section{Conclusions}
Seeking a weak lensing shear measurement methodology that can achieve systematic errors below 1 part in $10^3$, we examine the origin of the larger biases reported by VB and MBLB for measurement of noiseless galaxies using several current object-fitting methodologies.  We show that shape biases can be generated when fitting with the basis sets are (necessarily) truncated to finite size or with perfectly elliptical galaxy models.  A multiplicative shear bias will typically occur when the radial profile of the galaxy is not well matched by the finite basis set.  This could be remedied by choosing a finite basis that is complete over all galaxy shapes, but such a basis has not yet been devised.

One could extend the fitting functions to higher order to reduce the bias to acceptable levels---VB show this to work for most of their cases---however we demonstrate that there is a fundamental limitation to this approach.  If the ``roundness test'' being applied to the pre-seeing image extends to regions of $k$ space that have been highly or fully suppressed by the MTF of the image, then any attempt to fully eliminate the bias will cause the noise in the shape determination to diverge.  This will occur whenever one is trying to use a basis set that is size-matched to a pre-seeing galaxy image that is smaller than the PSF.  

We propose a new shape measurement, ``Fourier-Domain Null Tests'' (FDNT), which constructs linear functions of the galaxy's Fourier representation that are explicitly designed to maintain bounded noise in the presence of an MTF, and which are pure dipole and quadrupole functions of the pre-seeing galaxy intensity pattern.  Following the BJ02 and NB07 prescriptions for ``geometric'' shape measurements, the centroid and shear of the FDNT functions are varied until the tests are nulled, and the galaxy is assigned a shape according to the shear that nulled the tests.  This shape is guaranteed to transform in a well-known way under applied lensing shear.

No model fitting is required for the FDNT method if a full real-space observed image is available.  If the image has missing pixels or a restricted mask, then we can use basis-set-fitting methods on the {\em observed} image to interpolate the missing data and estimate the Fourier-domain image.  We demonstrate that the FDNT method obtains negligible multiplicative errors when applied to galaxies with pure elliptical isophotes, even for poorly resolved deVaucouleurs-profile galaxies with high ellipticity.  When using basis sets to estimate the Fourier transform, measurable biases return, depending upon the choice of basis set, but these biases are always smaller with FDNT tests than with the NB07 method.

We identify a second shear-measurement bias that is likely even more pervasive.  This one occurs when the measured ellipticity of the
underlying galaxy varies with the scale of weight function applied by the measurement.  When we measure the shape of such an ``ellipticity gradient'' galaxy we must be sure that the weighting function stays fixed if the galaxy is sheared.  In forward-fitting shape measurement methods, the effective weighting function arises implicitly from a combination of the fitted galaxy model, the noise distribution, and the PSF.  This implicit function {\em does} typically change as a galaxy is distorted, inducing a shear measurement bias.  Since our FDNT method controls the weighting function explicitly, we can determine and correct the ellipticity gradient effect on shear measurement.  We demonstrate that this reduces multiplicative biases by factors of 20--100 for test galaxies with ellipticity gradients.  The result is a shear measurement method yielding multiplicative biases in ring tests that in the worst case are at the target $10^{-3}$ level, and well below in most cases.

Both of the biases that we describe arise because truncated basis-set-fitting methods are weighting the pre-seeing images in a way that we do not explicitly control.  The FDNT method makes the weighting of our roundness tests explicit so we can avoid degraded regions of $k$ space and correct for ellipticity gradients.  One might be chagrined that the $EG$ corrections are an empirical calibration correction that must be measured on the (noisy) galaxy images, much like the polarizability factors of the commonly used KSB method \citep{KSB}.  Certainly noise on the estimate of $EG$ is an issue, since it enters non-linearly as $1/(1-EG)$ in the calculation of responsivity.
We note however that polarizability corrections are typically $O(0.1)$ or larger while the $EG$ correction is this large only for our ``torture test'' galaxies, and would more typically be 0.01--0.03.  Since the $EG$ correction are smaller to start with, they do not need to be determined to as high relative accuracy as the KSB polarizabilities.

We have investigated an idealized case in this paper, with well-sampled images having no missing pixels, no crowding,  uniform white noise, Gaussian PSFs, and very high $S/N$ on a single image.  Even in this case we note that obtaining $<10^{-3}$ multiplicative errors requires great care, {\it e.g.} the use of nonlinear shear responsivities as described in the Appendix, and knowledge of the PSF to very high accuracy.
In a future paper we will extend the FDNT method to finite noise.  We note that extension of the FDNT method to measurement of a galaxy imaged in multiple exposures is straightforward---one simply creates a null test that is the $S/N$-weighted sum of the null tests for the individual exposures.  We iterate the common centroid and shape of these tests until the summed test is nulled.  Each exposure can have its own PSF, and unlike fitting methods, we need not even assume that the galaxy has the same underlying appearance in each exposure.  So FDNT is easily applied to the full collection of multicolor images collected for a single galaxy.

This work is supported by grant AST-0607667 from the NSF and DOE grant
DE-FG02-95ER40893.  We thank the anonymous referee for his/her help in clarifying the paper.  We also thank Donnacha Kirk and Sree Balan for their substantial efforts in producing new realizations of the GREAT08 low-noise images for the blind tests reported herein.

\appendix
\section{Third-order shear responsivity}
Consider a population of $N$ perfectly elliptical galaxies with axis ratios $a/b=\tanh(\eta/2)$ and position angles uniformly distributed around $0<\beta<2\pi$.  If these are distorted by a lensing field with shear $\gamma$ such that they are magnified by $1+\gamma$ along the $x$ axis and demagnified by $1-\gamma$ on the $y$ axis, then a measurement of the lensed galaxies will yield ellipticities $\eta^\prime$ and position angles $\beta^\prime$.  We assume that we will infer the shear by calculating the mean of the ellipticity component $\eta_1$:
\begin{equation}
\langle \eta_1^\prime \rangle \equiv {1 \over N} \sum w(\eta^\prime) \eta^\prime \cos 2\beta^\prime.
\end{equation}
Here we allow for application of some weighting function $w(\eta^\prime)$ to the averaging, as suggested in BJ02 to reduce shot noise in the estimator.
We wish to determine how $\langle \eta_1^\prime \rangle$ depends upon the input shear $\gamma$, keeping terms to third order in $\gamma$.  The exact equations for shear addition are equations (2.12) of BJ02.  In our case they become
\begin{eqnarray}
\cosh \eta^\prime & = &  (1+2\gamma^2)\cosh \eta + 2\gamma(1+\gamma^2) 
\sinh \eta \cos 2\beta \nonumber \\
\sinh\eta^\prime \sin 2\beta^\prime & = & \sinh \eta \sin 2\beta
\label{addeta}
\end{eqnarray}
After solving these equations for $\eta_1$ in terms of $\gamma$ (to third order), $\eta,$ and $\beta$, and then averaging over $\beta$, we obtain
\begin{eqnarray}
\langle \eta_1^\prime \rangle & = & 2\gamma{\cal R}_1(\eta) + 2\gamma^3{\cal R}_3(\eta), \\
{\cal R}_1(\eta) & = & \eta w^\prime + w + Yw, \label{r1pure} \\
{\cal R}_3(\eta) & = {1 \over 4} \bigg\{  & 
  w\left[ 2-Y+ {Y^2(Y-1) \over \eta^2} \right]  \nonumber \\
 & & + 2\eta w^\prime \left[ 1 - {2Y(Y-1)\over \eta^2} + {3Y^2 \over 2\eta^2} \right] \nonumber \\
 & & +  w^{\prime\prime}(2Y+3) + \eta w^{\prime\prime\prime} \bigg\},  \\
Y & \equiv & {\eta \over \tanh\eta}.
\end{eqnarray}
When observing a population of galaxies, we will not know the unlensed $\eta$ values, so the responsivities ${\cal R}_1$ and ${\cal R}_3$ will need to be calculated at the observed (lensed) values $\eta^\prime$.  The relation of ${\cal R}_1$ as determined from the lensed distribution to the ${\cal R}_1$ at the unlensed $\eta$ is, to second order in $\gamma$, 
\begin{equation}
\sum {\cal R}_1(\eta^\prime) = N\left[{\cal R}_1(\eta) + 2\gamma^2 {\cal R}_3(\eta)\right].
\end{equation}
For the cubic term ${\cal R}_3$ we need only note that the mean value is the same, to linear order in $\gamma$, whether we average over lensed or unlensed $\eta$ values.

We assume now that we have a catalog of observed ellipticities $\eta_{1,i}$, for galaxies $i\in\{1,\ldots,N\}$, each assigned a weight $w_i$ and the linear and cubic responsivities ${\cal R}_{1,i}$ and ${\cal R}_{3,i}$ assigned from the observed ellipticity for each galaxy.  We define the linearized shear estimator
\begin{equation}
\hat \gamma_1 \equiv { \sum w_i \eta_{1,i} \over 2\sum {\cal R}_{1,i}},
\end{equation}
then the above equations imply that the expectation value of the linearized shear estimator will be related to the true reduced shear via
\begin{equation}
\left\langle \hat \gamma_1 \right\rangle = \gamma_1\left(1 - \gamma^2 { \sum {\cal R}_{3,i} \over \sum {\cal R}_{1,i}}\right).
\end{equation}
The same equation holds for the other shear component.
Given an linearized estimated shear $(\hat \gamma_1, \hat \gamma_2)$, one can quickly solve for the true shear $(\gamma_1,\gamma_2)$ that would produce this linearized estimate.  Alternatively the power spectra, etc., of the linearized estimator can be predicted from theoretical models for the power spectrum (and higher-order functions) of $\gamma$.  

We validate numerically that this procedure yields shear estimates accurate to $O(\gamma^5)$ for any shape measurements which transform under shear according to the exact shear addition formulae from BJ02.

In this work we show that most shape measurement methods have an ellipticity gradient correction $EG$ that alters the mean shear response along the principal axis.  The expression for ${\cal R}_1$ should be altered from \eqq{r1pure} as
\begin{equation}
{\cal R}_1(\eta)  =  {\eta w^\prime + w  \over 1- EG} + Yw.
\label{r1fixed}
\end{equation}
We have not calculated the effect of the ellipticity gradient on the cubic responsivity ${\cal R}_3$.
\end{document}